# Beyond Resilience: Antifragility in Critical Infrastructure Cybersecurity

Stephen Flowerday[1,*], Mauricio Papa[2] and Ethan Flowerday[3]

[1]*School of Computer and Cyber Sciences, Augusta University, Augusta, GA 30912, USA; sflowerday@augusta.edu*

[2]*Tandy School of Computer Science, The University of Tulsa, Tulsa, OK 74104, USA; mauricio-papa@utulsa.edu*

[3]*School of Cyber Studies, The University of Tulsa, Tulsa, OK 74104, USA; ethan-flowerday@utulsa.edu*

** Correspondence: sflowerday@augusta.edu*

**ABSTRACT**

Critical infrastructure cybersecurity increasingly needs frameworks that move beyond recovery toward bounded improvement under disruption, but empirically grounded theories for operational technology remain limited. This paper develops a Theory of Antifragility (AFT) for critical infrastructure (CI) cybersecurity, anchored in a five-state Resilient System Model and a bounded mathematical definition built around Jensen gain and post-disruption gain. A two-layer empirical design pairs a CI-relevant subset of the CISSM Cyber Events Database with the HAI hardware-in-the-loop industrial control dataset and tests three confirmatory hypotheses and one exploratory proposition. OT-adjacent sectors show significantly higher shares of disruptive or mixed events than comparison sectors (65.3% versus 46.8%, $p < 0.001$) and a heavier concentration of physical-attack and data-attack subtypes. In HAI, attack-labeled observations were 7.43 times more likely than normal observations to exceed the 95th percentile of baseline deviation ($p < 0.001$). Across successive attack windows, mean process-state deviation declined significantly (Spearman $\rho = -0.688$, $p = 0.007$), providing evidence of measurable response variation rather than proof of adaptive gain. Together, the findings establish two prerequisites for future antifragility testing: differentiated fragility burden and process-level perturbation observability.

## 1. INTRODUCTION

Cyber incidents in critical infrastructure matter because their effects extend well beyond the compromised organization. Disruption in one entity can ripple into energy delivery, transportation, public administration, health services, food supply, and public confidence. The Colonial Pipeline disruption and the JBS cyberattack remain clear examples of how a cyber event can quickly become a societal one [1,2]. Current guidance for operational technology and cyber-resilient systems reflects that reality by emphasizing continuity of function, safety, recovery, and adaptation under hostile conditions [3,4]. As information technology (IT) and operational technology (OT) continue to converge, sectors once treated as operationally separate inherit broader digital exposure and more complex interdependencies [3,4].

A conceptual problem remains. In cybersecurity scholarship, resilience, robustness, reliability, recovery, and adaptation are often treated as if they were interchangeable. They are not. Work in ecology, resilience engineering, and organization studies shows that a system can be stable without being resilient, and it can recover without becoming better [5–7]. In critical infrastructure, that distinction matters because a system may repeatedly return to service after comparable shocks and still remain vulnerable to the same pathways of failure.

Antifragility sharpens the distinction. In Taleb's formulation, fragile systems are disproportionately harmed by volatility, whereas antifragile systems benefit from bounded volatility because their response is convex rather than concave [8–10]. Johnson and Gheorghe extend this line of thought by treating fragility, robustness, and antifragility as positions on a measurable systems continuum [11]. That perspective is especially useful in cybersecurity, where persistent variation in adversary behavior makes recovery-only thinking increasingly inadequate.

Critical infrastructure cybersecurity therefore needs a theory of improvement, not only a theory of restoration. A resilient system bounces back; an antifragile system bounces forward. A related distinction matters as well: chaos engineering is the verb, while antifragility is the noun. Controlled perturbation can reveal weakness and create conditions for improvement, but it is not, by itself, evidence that a system has become antifragile [12–14]. Few empirically grounded frameworks explain how improvement through disruption should be defined, bounded, and tested in cyber-physical environments.

This study makes three contributions. First, it develops a Theory of Antifragility (AFT) for critical infrastructure cybersecurity, grounded in a five-state model and a bounded mathematical definition. Second, using cyber-event data, it shows that fragility burden can be differentiated empirically across critical-infrastructure sectors. Third, using industrial-control proxy data, it shows that bounded cyber-physical perturbations can be measured at the process level, a prerequisite for later antifragility testing in OT/ICS settings. The paper does not claim completed antifragile validation; it offers a foundation for future adaptive, self-learning, and safety-bounded cybersecurity research in critical infrastructure. In that sense, it extends Flowerday et al.'s conceptual framing of resilient and antifragile cybersecurity systems [15] by grounding the argument in formal definition and data.

The paper also addresses three questions that are often blurred together in the literature: where fragility is concentrated across critical infrastructure sectors, whether bounded cyber-physical perturbation can be observed and measured in OT/ICS process data, and what additional evidence would be required before a system could credibly be classified as antifragile. Section 3 translates those questions into three confirmatory hypotheses and one exploratory proposition. The distinction matters for current work on intelligent security systems. Adaptive and self-learning defense is a major research direction, but in critical infrastructure the real question is not simply whether a system can update after stress. It is whether it can do so safely, measurably, and in a way that leaves it better prepared for the next bounded disruption. The paper therefore asks whether the empirical conditions needed to test antifragility—measurable perturbation and structured system response—can be observed in public cyber-event and OT/ICS data.

## 2. THEORY AND RELATED LITERATURE

### 2.1 Resilience and Its Limits

Guidance on OT security and cyber resiliency rightly places performance, reliability, and safety alongside security [3,4]. Resilience, however, is not the same as improvement. Holling's distinction between resilience and stability, Woods's clarification of multiple resilience concepts, and Munoz et al.'s separation of resilience, robustness, and antifragility all point in the same direction: returning to an acceptable operating condition is not the same as becoming stronger because of adversity [5–7]. In practice, a system may recover repeatedly and still retain the same brittle dependencies, attack surface, and organizational blind spots.

That is why resilience, though essential, is not an adequate endpoint for critical-infrastructure cybersecurity. A restoration-centered posture can reduce downtime and still leave the system unchanged in any deeper sense. What it lacks is an explicit mechanism through which disruption becomes a source of durable gain. In OT/ICS settings, that gap is especially important because continuity requirements can conceal deeper fragility. A plant may continue to operate after a disruption, yet do so through manual overrides, temporary workarounds, or exhausted operator effort. Such a system is operationally persistent, but not structurally improved.

### 2.2 Antifragility as Bounce Forward

Taleb defines antifragility as the opposite of fragility: a property through which volatility, randomness, and bounded disorder can generate improvement rather than only harm [8]. The phrase bounce forward is useful because it makes the contrast with resilience immediately intelligible, but the concept becomes analytically meaningful only when linked to a payoff function. Taleb's short clarification in Nature and the fuller Taleb–Douady formulation do exactly that by tying antifragility to convex response under variability [9,10].

In engineering terms, this matters because it turns antifragility from metaphor into a property that can, at least in principle, be tested. Johnson and Gheorghe show that fragility, robustness, and antifragility can be placed on a measurable continuum for systems of systems [11]. More recent work on complex dynamical systems and system design reaches a similar conclusion: antifragility becomes meaningful when the response to stress can be modeled and when the region in which stress is beneficial is bounded [16,17].

This study adopts that bounded view. Antifragility in critical infrastructure cybersecurity is not treated as unlimited gain from chaos. It is defined as a safety-constrained systems property in which bounded disorder can lead to measurable post-disruption improvement. That boundedness matters because safety-critical systems cannot be allowed to "learn" through unrestricted failure. Any gain-from-disorder claim must therefore remain subordinate to process integrity, human safety, and operational continuity.

Prior work has clarified parts of the problem without closing it. Johnson and Gheorghe made fragility and antifragility more measurable at the systems-of-systems level [11]. Monperrus showed how controlled faults can support adaptive correction in software [14]. Flowerday et al. argued for resilient and antifragile cybersecurity under conditions of uncertainty [15]. What remains underdeveloped is a bounded formulation tied to both sector-level evidence and process-level OT/ICS telemetry. That is the narrower contribution of this paper. In engineering terms, a system is antifragile when its post-event configuration performs measurably better on the same bounded stressor, not merely when it survives.

### 2.3 The Resilient System Model

To avoid conceptual slippage, this study uses a five-state Resilient System Model. The model differentiates among fragile, reliable, robust, resilient, and antifragile systems, and makes explicit what is often only implied in resilience discussions [15]. Table 1 summarizes the five states. Fragile systems degrade disproportionately under bounded stress. Reliable systems perform well in familiar conditions but have a narrow adaptive range. Robust systems resist bounded perturbation without learning from it. Resilient systems recover to acceptable operation after disruption. Antifragile systems improve because of bounded disorder and therefore move beyond restoration alone.

**Table 1: Definitions of the five system states**

| State | Core response | Practical meaning | Indicative payoff shape |
|---|---|---|---|
| Fragile | Disproportionate degradation | Small disturbances create outsized harm | Concave |
| Reliable | Stable in familiar conditions | Performs well inside a narrow envelope | Bounded linear |
| Robust | Resistance without learning | Holds function under bounded stress | Near linear |
| Resilient | Recovery to acceptable operation | Returns to service after disruption | Bounded recovery |
| Antifragile | Improvement because of disruption | Learns and performs better after bounded stress | Convex |

The model matters in two ways. Conceptually, it keeps robustness, resilience, and antifragility from collapsing into one another. Methodologically, it creates a bridge from theory to data. The question is no longer just whether a system survived, but what kind of state transition the disturbance produced.

The reliable category is especially important in cybersecurity, where many systems appear strong only because they operate inside a narrow envelope of expected conditions. Reliability under familiar load, familiar traffic, or familiar adversary behavior is not the same as antifragility. It may not even qualify as resilience. The model therefore treats antifragility as the farthest point on a progression, not as a rhetorical synonym for "very resilient."

One of the clearest ways to distinguish resilience from antifragility is to compare systemic growth with sacrificial robustness. An ablative heat shield survives extreme thermal stress by consuming itself. Explosive reactive armor protects by detonating outward and sacrificing a module of the system. Both preserve the larger asset, but neither becomes stronger because of the event. Both require restoration afterward.

The same distinction applies in cybersecurity. A system that absorbs a ransomware event, restores service, and remains exposed to the same logic path or dependency structure is resilient at best. Antifragility would require the system to convert the event into a durable reduction in future susceptibility or an increase in safe adaptive range. That could take the form of better fallback logic, lower coupling, automated reconfiguration, faster recovery, or more effective detection of the same perturbation family. Without that improvement step, the system has survived disorder, but it has not gained from it.

The difference between surviving by expenditure and surviving by learning is central to the present argument. Critical infrastructure already contains many examples of sacrificial robustness: manual bypass, segmented fallback, or capacity reserve. Those practices are valuable, but they are not the same as antifragility unless the disturbance is translated into lasting architectural advantage.

### 2.4 Chaos Engineering as the Verb; Antifragility as the Noun

Chaos engineering is widely understood as disciplined experimentation on a system to build confidence in its ability to withstand turbulent conditions [12,13]. The canonical example is Netflix's Simian Army, which used controlled failure injection to expose hidden dependencies in production-like environments and force more resilient service design [12,13]. That definition is useful, but it should not be confused with antifragility itself. Chaos engineering is the verb. Antifragility is the noun.

Monperrus makes a similar point in the software context: systems do not become antifragile simply because faults are injected into them; they become antifragile when exposure to controlled faults leads to adaptive fault tolerance, runtime correction, or other durable improvements [14]. The distinction is especially important in OT/ICS environments, where perturbation must remain bounded by safety and process integrity. Stress alone is not evidence of gain. It is only an instrument.

This noun-versus-verb distinction resolves a common ambiguity in cybersecurity writing. A system may practice chaos engineering while remaining merely robust or resilient. Conversely, a system might display local antifragile behavior after an event without ever having been subjected to a formal chaos-engineering program. The theory therefore separates mechanism from state. Controlled perturbation is one pathway to antifragility, not a definition of it.

**2.5 Mathematical Foundations**

At the center of the theory is a utility or performance function that responds convexly to bounded disorder. Let U(θ, s) denote the utility of system configuration θ under a bounded perturbation s drawn from a probability distribution S. Throughout Equations (1)–(3), θ denotes the same original system configuration, s denotes a realized bounded stressor in Equations (1) and (3), and S denotes the bounded stressor distribution used in the expectation in Equation (2). Local antifragility requires a positive second derivative over an admissible stress region, as shown in Equation (1). A corresponding Jensen-gain criterion is given in Equation (2).

$$\frac{\partial^2 U(\theta, s)}{\partial s^2} > 0 \quad (1)$$

$$J(\theta, S) = \mathbb{E}[U(\theta, S)] - U(\theta, \mathbb{E}[S]) \quad (2)$$

Here, E denotes the expectation operator. When J(θ, S) > 0, bounded volatility is beneficial on average; when J(θ, S) < 0, the response remains fragile [9,10,16].

The notation is intentionally general as this is a measurement framework for personnel to adapt for the classification of their system. The perturbation s may be represented as a scalar stress intensity, a multivariate vector of attack or process features, or a time-dependent perturbation path. In OT/ICS settings, the specific representation depends on the deployment environment, available telemetry, and system constraints. The empirical HAI example in this paper uses multichannel process data and summarizes perturbation magnitude through z-score, RMS, and Mahalanobis-distance measures, with the latter accounting for covariance among correlated sensor channels.

Convex response alone, however, is not sufficient. A system may display local convexity and still fail to improve after disruption. For that reason, the theory adds a second requirement. Let Φ denote an adaptation operator that updates the system after exposure to bounded perturbation. Post-disruption gain, Gp, is defined in Equation (3).

$$G_p(\theta, s) = U(\Phi(\theta, s), s) - U(\theta, s) \quad (3)$$

When Gp > 0, the adapted configuration performs better than the original configuration under the same bounded perturbation. This keeps antifragility from collapsing back into resilience. A resilient system returns to baseline; an antifragile system improves relative to baseline. In practical OT/ICS settings, the adaptation operator could take the form of controller retuning, segmented reconfiguration, rule refinement, or operator-guided automation, provided the change remains within the safety envelope. The intelligent-systems framing becomes concrete at this point: a self-learning detector or policy updater matters only if its update can be tied to a bounded perturbation and a measurably better response on retest. Equations (1) and (2) capture gain from variability; Equation (3) requires that disruption leave the system in a better state than before.

Equation (3) is intended as a practical before/after measurement criterion rather than a fitting or optimization procedure. More precisely, the adaptation operator is not tuned to the specific realization that prompted it, so the criterion measures generalization within the stress class rather than fit to a single episode. In operational OT/ICS settings, adaptation is typically a documented human-in-the-loop engineering change (with multiple constraints such as cost, security, deployability, operational continuity, and time) rather than an automated parameter update. The purpose of Equation (3) is therefore to measure whether the post-adaptation configuration performs better than the pre-adaptation configuration under a controlled perturbation. In practice, evaluating different perturbations against both the original and adapted configurations is often infeasible because only one physical system is available for measurement comparison, unless a digital twin, reversible configuration, or separate control system exists.

Taleb and Douady's H-heuristic sharpens the argument further by treating antifragility as a curvature problem and asking whether performance becomes more favorable as bounded dispersion increases [10]. In practical terms, the heuristic matters because it frames antifragility as positive sensitivity to variability rather than vague post hoc adaptation. The present study does not estimate the H-heuristic directly from the CISSM subset, but the logic remains central: a persuasive antifragility claim must show more than survival. It must show that bounded disorder alters the response surface in a favorable direction.

Two implications follow. First, antifragility is necessarily comparative: there must be a before and an after. Second, antifragility is domain bounded: the same configuration may be antifragile with respect to one perturbation family and fragile with respect to another. That is not a defect in the theory. It is part of what makes it empirically credible.

Antifragile classification therefore requires simultaneous positivity of Jensen gain and post-disruption gain over a bounded disorder domain. This is a formal criterion rather than a conventional regression hypothesis. A system should be classified as antifragile only when bounded volatility is beneficial on average and the post-disruption configuration performs better than the pre-disruption configuration.

To operationalize the framework, the paper defines four measurable components. In addition to Jensen gain (Equation (2)) and post-disruption gain (Equation (3)), it introduces susceptibility reduction (Equation (4)) and recovery improvement (Equation (5)). In Equations (4) and (5), the before/after labels refer to susceptibility or recovery time before and after the adaptation step; superscript 1 denotes the initial perturbation episode, and superscript k denotes a later comparable retest episode. The components can be combined into a normalized Antifragility Score (Equation (6)), with weights that sum to one.

$$\Delta S = S_{\text{before}} - S_{\text{after}} \tag{4}$$

$$\Delta T_{\text{rec}} = T_{\text{rec}}^{(1)} - T_{\text{rec}}^{(k)} \tag{5}$$

$$\text{AFS} = w_1 J_{\text{norm}} + w_2 G_{\text{norm}} + w_3 S_{\text{norm}} + w_4 T_{\text{norm}} \tag{6}$$

The present study does not estimate all four Antifragility Score components from the available data. Instead, it tests two prerequisites for future score estimation: sector-level fragility differentiation and process-level perturbation observability. Equations (4) through (6) define the fuller framework for future adaptation-and-retest work. In plain language, the score asks four questions: does bounded variability help, does the post-event configuration perform better, is the system less susceptible next time, and does it recover faster?

### 2.6 Worked Simulation Example

To make the framework concrete, consider a simple synthetic utility curve measured under a bounded stressor ε following a uniform distribution on [−0.5, 0.5]. Suppose the pre-adaptation response is $U_0(\varepsilon) = 1 - 0.3\varepsilon^2$, which is locally concave and therefore fragile. In that case, the Jensen-gain criterion in Equation (2) is negative: $E[U_0(\varepsilon)] = 0.975$, while $U_0(E[\varepsilon]) = 1.000$, yielding a Jensen gain of −0.025. After a bounded adaptation step, let the response become $U_1(\varepsilon) = 1 + 0.2\varepsilon^2$ over the same admissible domain. The Jensen gain then turns positive, with a value of 0.017. Furthermore, post-disruption gain in Equation (3) is non-negative for a held-out test stress: $Gp = (1 + 0.2\varepsilon^2) - (1 - 0.3\varepsilon^2) = 0.5\varepsilon^2$.

If susceptibility and recovery-time deltas are non-negative after the same perturbation family is replayed, the normalized Antifragility Score in Equation (6) becomes positive. The values used here are illustrative rather than empirical; their purpose is to show how the framework would be used in practice. The corresponding workflow is:

1. Establish baseline telemetry under normal operating conditions.
2. Define the bounded perturbation family.
3. Measure baseline susceptibility and recovery characteristics.
4. Apply perturbation.
5. Document the adaptation operator.
6. Replay a comparable perturbation.
7. Recompute Jensen gain, post-disruption gain, susceptibility reduction, and recovery improvement.
8. Compute the Antifragility Score to classify the system state.

## 3. HYPOTHESIS DEVELOPMENT

The broader AFT rests on six constructs: Robustness, Recoverability, Learning and Feedback Integration, Adaptive Capacity, Uncertainty Orientation, and Redundancy and Optionality. For the present paper, those constructs are narrowed into three confirmatory hypotheses and one exploratory proposition that fit the available public data. H1 and H2 operationalize sector-level robustness and redundancy/optionality; H3 and H4 address process-level adaptive capacity and learning/feedback integration. Recoverability and Uncertainty Orientation remain in the framework, but the current public datasets do not support a clean direct test of them.

**H1. OT-adjacent critical-infrastructure sectors will exhibit a higher share of disruptive or mixed cyber events than non-OT-adjacent sectors.**

The rationale is straightforward: sectors that depend more heavily on physical process control should carry a larger share of events whose primary effects are service disruption, control impairment, or mixed operational damage rather than pure data theft.

**H2. OT-adjacent critical-infrastructure sectors will show a disproportionate concentration of attack subtypes with direct operational consequences, specifically physical attack and data attack, relative to non-OT-adjacent sectors.**

Utilities, transportation, and manufacturing are more exposed to cyber events that affect process state, industrial data integrity, or physical service continuity. Those sectors should therefore show higher frequencies of physical-attack and data-attack effects than sectors whose exposure is mainly information-centric.

**H3. In OT/ICS proxy data, attack-labeled windows will produce larger process-state deviations from baseline than normal windows.**

This hypothesis does not claim antifragility. It tests a necessary condition for future antifragility validation: that bounded cyber-physical perturbation is empirically measurable in industrial process data.

**H4. In OT/ICS proxy data, deviation magnitude can be measured across successive attack windows, providing exploratory evidence of variation across repeated bounded perturbations.**

Process-state responses across successive attack windows may vary systematically, providing exploratory evidence on response dynamics under repeated perturbation. This proposition does not assume system adaptation; it asks whether repeated bounded perturbation produces an observable change in process response, a necessary precondition for later adaptation-and-retest work under the formal criteria established in Section 2.5.

## 4. METHODOLOGY

### 4.1 Research Design

The empirical design has two layers. The first is macro and sectoral: it draws on the CISSM Cyber Events Database to evaluate fragility burden across selected critical-infrastructure sectors. The second is micro and process-based: it uses HAI 23.05 proxy data to examine whether attack windows generate measurable process-state deviation in OT/ICS telemetry.

The design does not claim completed field experimentation or a live adaptation-and-retest cycle. Instead, it tests two prerequisites for future antifragility validation: whether fragility can be differentiated empirically across sectors, and whether bounded cyber-physical perturbation can be measured at the process level.

Because live access to operational critical-infrastructure sites was unavailable, the study uses a proxy-based strategy built on public cyber-event and industrial-control data. This choice is both practical and conceptually important: theory without measurable proxy layers is difficult to carry forward in cybersecurity research. These two layers correspond directly to the theoretical requirements of bounded perturbation and observable system response outlined in Section 2.5.

The CISSM Cyber Events Database supports macro-level validation because it classifies cyber events by actor type, motive, industry, and event effect [18,19]. Its categorical structure supports the sector-level comparisons central to H1 and H2. The HAI 23.05 dataset supports the micro-level analysis because it captures industrial-control behavior from a hardware-in-the-loop testbed with attack-labeled intervals [20,21]. Together, the datasets support a layered argument: CISSM helps identify how fragility is distributed across CI-relevant sectors, while HAI shows whether bounded cyber-physical perturbation can be observed at the process level.

Sectoral cyber-event data capture the macro distribution of disruption but cannot reveal process-level response. Process telemetry, by contrast, captures system behavior under attack but does not show how fragility is distributed across sectors. The two layers are therefore complementary: CISSM maps where fragility concentrates, while HAI shows how bounded perturbation manifests in cyber-physical behavior.

### 4.2 CISSM Data and Critical-Infrastructure Subset

The CISSM codebook describes the database as a structured corpus of publicly available cyber events beginning in January 2014 and extending through March 2026 at the time of this analysis [19]. The full CSV used here contains 16,729 events. For this study, the analysis was limited to seven critical-infrastructure-relevant sectors: Utilities, Transportation, Information, Finance, Healthcare, Manufacturing, and Public Administration. Manufacturing combines NAICS sectors 31–33. Public Administration corresponds to NAICS Sector 92; where the CISSM file provided closely related administrative or public-service labels, those records were harmonized under Public Administration. Transportation was mapped to the CISSM transportation category, corresponding broadly to NAICS 48–49. The resulting seven-sector subset contains 10,589 events, as summarized in Table 2.

**Table 2: CISSM event counts and within-sector shares**

| Sector | Total events | Disruptive or mixed share (%) | Physical or data share (%) | OT-adjacent |
|---|---|---|---|---|
| Utilities | 321 | 67.9 | 47.7 | Yes |
| Transportation | 532 | 66.9 | 29.3 | Yes |
| Manufacturing | 785 | 63.1 | 46.0 | Yes |
| Public Administration | 3218 | 57.1 | 21.6 | No |

| Sector | Total events | Disruptive or mixed share (%) | Physical or data share (%) | OT-adjacent |
|---|---|---|---|---|
| Information | 1851 | 50.1 | 14.8 | No |
| Healthcare | 2266 | 38.9 | 32.5 | No |
| Finance | 1616 | 33.3 | 12.9 | No |

The subset was selected for two reasons. First, these sectors combine clearly critical service provision with enough event volume to support comparison. Second, they include both OT-adjacent sectors and sectors whose exposure is more information-centric, allowing a theoretically meaningful contrast.

For hypothesis testing, sectors were divided into two groups. OT-adjacent sectors were defined as Utilities, Manufacturing, and Transportation because these environments depend more directly on process control, industrial automation, and service continuity in cyber-physical settings [3,4]. The comparison group comprised Finance, Healthcare, Information, and Public Administration. Differences in event-type composition between the two groups were evaluated using Pearson chi-square tests, which assess whether observed frequencies differ significantly from expected frequencies under independence, together with Cramér's V, which provides a normalized effect-size measure scaled between 0 and 1. The grouping is theory-driven rather than causal: it captures a meaningful OT-versus-non-OT contrast without implying that sector membership alone explains the observed differences.

Event volume alone does not capture sectoral fragility burden. Public Administration carries the heaviest event load by far, yet OT-adjacent sectors show a consistently higher concentration of disruptive-or-mixed events, with Utilities, Transportation, and Manufacturing all exceeding 60% compared with a maximum of 57% in the non-OT group. The pattern matters theoretically: sectors more tightly coupled to physical process control appear to attract a qualitatively different threat profile, not merely a larger one. This distinction motivates the group comparison in Section 5 and aligns with H1. Public Administration's relatively high disruptive-or-mixed share is noted as a boundary case, possibly reflecting the operational continuity demands of government services.

To connect the macro data more directly to the theory of cyber-physical fragility, the event-subtype field was parsed into interpretable categories, including physical attack, data attack, internal and external denial of service, application-server exploitation, end-host exploitation, and message manipulation [19]. This allows comparison not only across broad event types but also across operationally relevant effects. The aggregation of physical and data attacks used in H2 is deliberate: physical attacks represent direct cyber-physical disruption, while data attacks can impair process control, degrade system visibility, or force manual intervention without immediate physical damage.

To identify which specific attack subtypes are disproportionately concentrated in OT-adjacent sectors, the event-subtype field was parsed by splitting compound entries into individual subtype flags. A chi-square test was then conducted for each subtype to assess whether its frequency differed significantly between OT-adjacent and non-OT groups, with Cramér's V reported as an effect-size measure.

Taken together, these distinctions show that fragility burden is not captured by event volume alone. OT-adjacent sectors display a more operationally concentrated threat profile, characterized by higher shares of disruptive, physical, and data-integrity effects, while more information-centric sectors show a greater emphasis on exploitative activity. This provides the empirical basis for the sector-level comparisons in Section 5.

**4.3 HAI Proxy Preprocessing**

The HAI 23.05 dataset captures coupled process behavior from a hardware-in-the-loop industrial-control testbed under both normal and attack conditions [20,21]. This analysis uses 65 informative process-variable channels across 54,000 test observations, of which 2,981 are attack-labeled (5.5%). Channels were retained only when their variance exceeded $1 \times 10^{-20}$ in the training set. This threshold excludes one numerical-artifact channel with effectively zero variance and avoids unstable standardization.

Baseline means and standard deviations were estimated from the normal operating data. For each informative channel, a standardized deviation score was computed from the test data using Equation (7).

$$z_{i,j} = \frac{x_{i,j} - \mu_j}{\sigma_j} \quad (7)$$

Here, $x_{i,j}$ is the observed value for row i and channel j, $\mu_j$ is the channel mean, and $\sigma_j$ is the corresponding standard deviation. The resulting $z_{i,j}$ score captures how far a sensor reading deviates from the normal operating baseline.

Two row-level summary measures were then derived: mean absolute z-score across all informative channels and a multivariate baseline-distance statistic computed as the root mean square of standardized deviations. Both summarize how far a given observation departs from the normal process state. Because ICS process variables are physically coupled, univariate deviation measures may overstate or understate perturbation magnitude when channels move jointly. Mahalanobis distance accounts for covariance structure and therefore provides a multivariate robustness check on whether attack-normal separation persists after correlated variation is incorporated. To test the robustness of the main separation, the study also computed a covariance-aware Mahalanobis distance on the same 65-channel subset using a regularized covariance estimate from the training partition. The Mahalanobis results are reported as a sensitivity check in Section 5.2. The z-score/RMS formulation remains the baseline because it is transparent, reproducible, and adequate for the paper's narrower aim of establishing perturbation observability.

The three measures used in the analysis capture different aspects of process-state deviation. Mean absolute z-score summarizes average displacement across channels and is relatively robust to noise in any single variable. RMS baseline distance puts more weight on larger departures, making it more sensitive to attack-driven excursions. Mahalanobis distance adds a covariance-aware check by accounting for correlated channel movement. The main analysis reports the transparent z-score and RMS measures; Mahalanobis distance is retained as a sensitivity check.

The difference between attack-labeled and normal observations was evaluated with a one-sided Mann–Whitney test, which asks whether attack-labeled observations tend to produce systematically larger baseline distances than normal observations without assuming normality. To make the result easier to interpret, the study also compared the proportion of attack-labeled and normal observations that exceeded the 95th percentile of the normal baseline-distance distribution.

To examine whether perturbation exhibits structure across repeated attack intervals, contiguous attack windows were identified and summarized using window-level statistics, including duration, mean deviation, peak deviation, and simple trend measures. These statistics do not constitute a test of adaptation, but they do provide descriptive evidence on whether system response varies across repeated bounded perturbations.

Although directionally expected, the result is important. The analysis shows that process-state deviation can be represented as a scalar quantity in publicly available ICS proxy data, satisfying the empirical prerequisite established in H3. The window-level analysis supports discussion of response variation across repeated perturbations, but it is exploratory and does not establish adaptive gain.

## 5. RESULTS

### 5.1 CISSM Analysis

The CISSM analysis uses sectoral variation in cyber-event profiles to connect sector-level fragility patterns with process-level OT/ICS perturbation evidence. The OT-adjacent grouping differs from the non-OT comparison group on each core burden measure reported in Table 3. For H1, OT-adjacent sectors show a higher share of disruptive-or-mixed events than non-OT sectors, 65.3% versus 46.8% ($\chi^2 = 190$, $p < 0.001$, Cramér's V = 0.134).

For H2, OT-adjacent sectors also had a substantially higher share of physical-or-data attack effects than the comparison sectors, 40.9% versus 21.4% ($\chi^2 = 286$, $p < 0.001$, Cramér's V = 0.164). The contrast becomes even sharper for physical attack alone: OT-adjacent sectors showed a physical-attack share of 4.2%, compared with 0.3% in the non-OT group ($\chi^2 = 251$, $p < 0.001$, Cramér's V = 0.154).

**Table 3: Event-level shares and grouped effect sizes for OT-adjacent versus non-OT sectors**

| Measure | OT-adjacent | Non-OT | $\chi^2$ | p-value | Cramér's V |
|---|---|---|---|---|---|
| Disruptive or Mixed Share (%) | 65.3 | 46.8 | 190 | <0.001 | 0.134 |
| Physical or Data Share (%) | 40.9 | 21.4 | 286 | <0.001 | 0.164 |
| Physical Share (%) | 4.2 | 0.3 | 251 | <0.001 | 0.154 |

The subtype analysis follows the same pattern. Events with a physical-attack component are far more common in OT-adjacent sectors (4.2% versus 0.3%). Events with a data-attack component are also more common (36.9% versus 21.1%). Application-server and end-host exploitation, by contrast, are more prevalent in non-OT sectors, consistent with their more information-centric exposure profile.

The sector detail is also informative. Utilities had the highest physical-attack share (14.0%) and the highest combined physical-or-data share (47.7%). Manufacturing showed the highest share of events containing data-attack effects (45.5%). Transportation had the highest disruptive-or-mixed share after Utilities (66.9%).

These results support H1 and H2. OT-adjacent sectors do not simply experience a different number of cyber events; they experience a different mix of events. The effect sizes are modest but meaningful in a large public event dataset, and they indicate that operationally exposed sectors carry a heavier disruption-oriented burden. The results therefore support the first empirical prerequisite for an antifragility framework: the distribution of fragility is structured and measurable rather than random or uniform.

**5.2 HAI Proxy Evidence**

The HAI proxy results support H3. In the test data, the 2,981 attack-labeled observations occurred in 14 distinct attack windows, with durations ranging from 56 to 628 seconds at a one-second sampling rate. The average window length was 212.9 seconds, with incidents persisting long enough to observe process-state effects.

Table 4 summarizes process-state deviation under normal and attack conditions. Using the 65 informative channels as features, the mean absolute z-score is 2.23 during attack-labeled observations and 1.53 during normal conditions, a ratio of 1.46. The RMS baseline-distance measure shows stronger separation, with mean values of 7.52 under attack and 3.92 under normal conditions, a ratio of 1.92. This stronger separation suggests that attack periods are characterized not only by elevated average deviation but also by more extreme departures in a subset of channels.

Tail behavior provides the clearest contrast. Attack-labeled observations are 7.43 times more likely than normal observations to exceed the 95th percentile of the normal baseline-distance distribution. A one-sided Mann– Whitney test on the RMS baseline-distance statistic confirms that attack-labeled observations exhibit significantly higher deviation than normal observations ($p < 0.001$).

A covariance-aware sensitivity check produced the same ordering. Mean Mahalanobis distance was 6.89 for normal observations and 18.57 for attack-labeled observations, a ratio of 2.69. Using the 95th percentile of the normal Mahalanobis distribution as a threshold, 79.4% of attack observations exceeded it, equating to a ratio of 15.88. The corresponding Mann–Whitney test also remained significant ($p < 0.001$). Thus, the separation between normal and attack conditions becomes more pronounced when using a covariance-aware multivariate check.

A plausible explanation for the larger deviation under attack is that HAI process variables are physically coupled through control-loop dynamics. Attacks that manipulate set points, actuator states, or control logic can propagate across multiple channels rather than appearing as isolated sensor departures. This interpretation is consistent with the stronger separation observed in RMS and Mahalanobis measures, both of which emphasize multichannel deviation.

**Table 4: HAI process-state deviation under normal and attack conditions**

| Measure | Normal | Attack | Ratio |
|---|---|---|---|
| Mean absolute z-score | 1.53 | 2.23 | 1.46 |
| RMS baseline distance | 3.92 | 7.52 | 1.92 |
| 95th percentile exceedance (%) | 5.00 | 37.10 | 7.43 |
| Mean Mahalanobis distance | 6.89 | 18.57 | 2.69 |
| Mahalanobis exceedance (%) | 5.00 | 79.4 | 15.88 |

Figure 1 plots mean process-state deviation across successive attack windows. The declining trend is statistically significant (Spearman $\rho = -0.688$, $p = 0.007$), which indicates that measured response varies across repeated perturbations. However, the pattern is not accompanied by consistent improvement relative to baseline conditions, and it cannot be attributed unambiguously to adaptation without attack-type labels. Accordingly, the result is treated as exploratory evidence of measurable response variation rather than proof of improvement. The monotone form of the decline is consistent with the kind of signal an improving response might produce under comparable attack conditions, but in this dataset it remains hypothesis-generating only.

Together, these results show that process-state deviation is observable and structured under bounded perturbation, satisfying the empirical prerequisite for antifragility testing established in H3. The observed variation across attack windows indicates measurable response dynamics, but it does not justify a claim of antifragility. That claim would require an explicit adaptation step and a comparable retest showing improvement in the formal components defined in Equations (3) through (6).

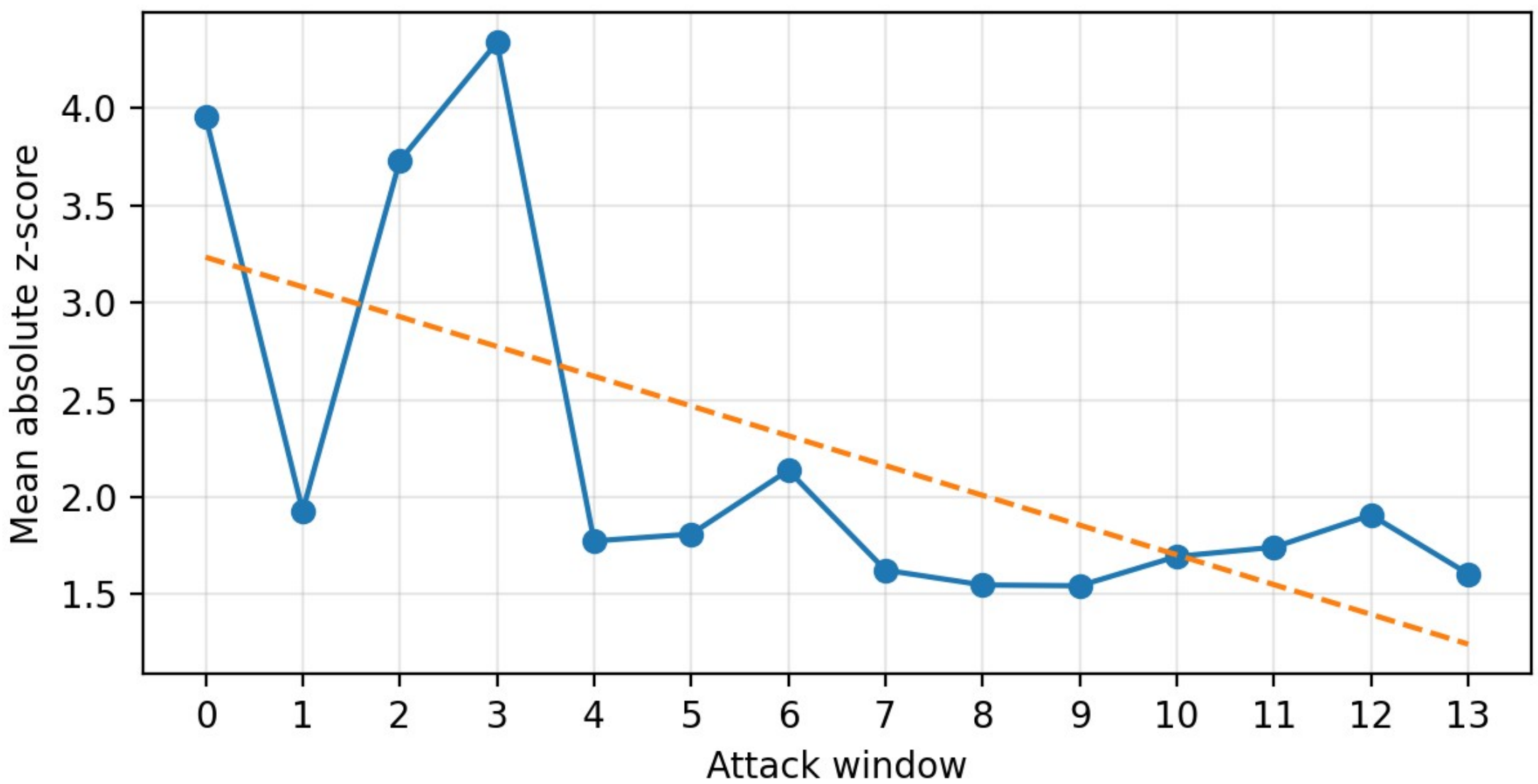

**Figure 1: Mean absolute z-score across successive attack windows (0-13) in HAI test data**

The declining trend is reported as exploratory evidence of response variation across heterogeneous attack windows; because attack-type labels are unavailable, it is not interpreted as evidence of adaptation.

## 6. DISCUSSION

The two empirical layers establish measurable prerequisites for antifragility testing in critical infrastructure cybersecurity. The CISSM data show that fragility burden is structured across critical-infrastructure sectors, while the HAI proxy data show that bounded perturbation is observable in industrial process telemetry. Together, they move the discussion beyond metaphor without overstating what the current data can prove.

That limitation is informative rather than fatal. The findings show that the prerequisites for antifragility testing are present and measurable, while also clarifying where current evidence ends. The contribution of the paper is therefore not to claim antifragility, but to establish a credible path for testing it.

The CISSM results show that fragility burden is not uniform across sectors. OT-adjacent sectors differ not only in event volume but in event composition, with higher shares of disruptive, physical, and data-integrity effects. That distinction matters because antifragility theory assumes heterogeneous exposure to stress. At the same time, the effect sizes remain modest (Cramér's V = 0.134–0.164), so the results should be read as structured differences in a public event corpus rather than as large causal separations between sector types.

The subtype analysis sharpens the point. Physical attack is concentrated in OT-adjacent sectors by a factor of roughly 14, while application-server and end-host exploitation dominate the non-OT group. That distinction matters theoretically because physical and data-integrity attacks directly threaten process continuity in ways that recovery-centered frameworks are not designed to address. They are precisely the attack vectors for which bounce-forward improvement, rather than bounce-back recovery, would represent a meaningful advance.

The HAI results complement this at the process level. They show that cyber-physical perturbations produce measurable and localized deviations relative to a stable baseline. That is a necessary condition for antifragility testing: without observable displacement, there is no basis for measuring improvement. The results also show that perturbations are structured rather than random, and that system response varies across repeated attack windows. However, this variation does not constitute evidence of improvement, because post-disruption behavior does not consistently converge toward baseline and the attack windows are heterogeneous.

These results clarify the relationship between resilience and antifragility in critical infrastructure. Resilience remains the operational baseline, ensuring safety, recovery, and continuity. The present findings suggest that resilience should be treated as a floor rather than a ceiling. The sectors most exposed to operationally disruptive effects are precisely those for which recovery alone may be insufficient.

The sector-level differences are instructive in this regard. Public Administration carries the largest event burden, but OT-adjacent sectors exhibit a more operationally concentrated threat profile. Improvement pathways are therefore unlikely to be uniform. Sectors dominated by exploitative events may improve through better identity management and recovery processes, while sectors exposed to physical and data-integrity attacks require stronger process-aware monitoring, segmentation, control validation, and safe fallback design.

The results also reinforce the importance of distinguishing between perturbation and improvement. The HAI analysis shows that bounded perturbation reveals system behavior, but it does not generate gain on its own. Demonstrating antifragility requires an explicit adaptation step followed by a repeat test under comparable conditions. This distinction aligns with the separation between chaos engineering as a method and antifragility as a system property.

A practical path for future work follows from this structure. A valid antifragility test in this setting requires three elements: (i) a bounded and repeatable perturbation, (ii) an explicit adaptation step, and (iii) a rerun demonstrating measurable improvement, such as reduced susceptibility or faster recovery under the same stress class. The present study establishes the first element and provides evidence for observability, while leaving the adaptation-and-retest requirement for future work.

The results also speak to the broader discussion of intelligent and adaptive cybersecurity systems. Adaptive behavior does not automatically imply improvement. Systems that update in response to disruption may shift risk, overfit to specific events, or reduce safety margins. The framework proposed here requires that any claimed improvement remain bounded, interpretable, and safe. That constraint is essential in critical infrastructure, where the cost of incorrect adaptation can exceed the cost of delayed improvement.

### 6.1 Practical Implications for Operators

For operators and practitioners, the framework suggests four near-term steps that do not require a full adaptation-and-retest program. First, establish stable baseline statistics for channels most exposed to cyber-physical disruption. Second, treat post-incident change as a documented adaptation step rather than an informal recovery tweak. Third, use bounded perturbation windows where the safety case permits. Fourth, rerun comparable perturbations after change and measure whether susceptibility, recovery time, or process deviation has improved. These practices do not require operators to induce unsafe failure. They require disciplined measurement of bounded stress and explicit evidence that the post-event configuration performs better.

## 7. LIMITATIONS AND FUTURE RESEARCH

The clearest limitation is that the study does not observe a completed adaptation cycle. CISSM provides rich coded event structure but sparse direct harm fields, which is why a transparent proxy was required. HAI provides attack labels and process telemetry, but not an explicit post-adaptation rerun. The present data can therefore distinguish fragility burden and process-state deviation, but they cannot directly estimate all Antifragility Score components.

Scale is the second limitation. CISSM is sectoral and organizationally heterogeneous, while HAI is process-level and testbed-based. Together they form a strong layered design, but they do not close the gap between macro fragility mapping and micro antifragility proof. The paper also proposes a formal antifragility criterion without estimating all score components from the available data. For that reason, the present study should be read as a rigorous foundation rather than as empirical closure.

### 7.1 CISSM Measurement Biases

Public-event reporting bias and coder judgment in CISSM also deserve explicit acknowledgment. Operationally visible OT incidents may be more likely to be recorded than smaller events in non-OT settings, and several coded fields, especially motive and event type, still depend on human interpretation of open-source reports. Broad NAICS aggregation can further compress within-sector variation. These limitations do not invalidate the observed sectoral patterns, but they mean the results should be interpreted as structured evidence of differential fragility burden rather than as precise incidence rates.

### 7.2 Future Directions

Future work should proceed in four directions. First, the CISSM design should be extended with richer organization-level or incident-case data that record post-event control changes. Second, HAI should be paired with an explicit adaptation operator so that the same perturbation family can be replayed after change. Third, the proxy layer should be extended using vulnerability priors informed by the CISA Known Exploited Vulnerabilities catalog and MITRE ATT&CK for ICS [22–24]. Fourth, related public OT/ICS datasets, including ELECTRON DNP3 and ICS-ADD, should be used to test whether perturbation observability generalizes beyond HAI [25,26]. Recent work on antifragile critical infrastructure and resilience metrics provides a useful basis for this next stage [27,28].

## 8. CONCLUSION

Critical infrastructure cybersecurity cannot stop at recovery. Systems that merely return to service remain exposed to the same future disorder unless additional steps are taken. This paper therefore argues for a move beyond resilience toward antifragility: a bounded, safety-constrained capacity for improvement under disruption.

The paper makes three contributions. First, it offers a theoretical contribution by defining antifragility for critical infrastructure cybersecurity as bounded improvement under disorder and anchoring that definition in a five-state Resilient System Model. Second, it makes a methodological contribution by showing how antifragility can be examined through a layered empirical design that combines sector-level event evidence with process-level telemetry. Third, it provides empirical evidence that two prerequisites for future antifragility testing—fragility differentiation and perturbation observability—are measurable in public data.

The empirical results support this structure. The CISSM analysis shows that fragility burden is differentiated across sectors, with OT-adjacent environments exhibiting more operationally disruptive event profiles. The HAI analysis shows that bounded cyber-physical perturbation is observable and measurable in process telemetry. Together, these findings establish the empirical prerequisites for antifragility testing without claiming completed antifragility.

The current data do not demonstrate that any system is antifragile. They show that fragility can be differentiated and that perturbation can be measured in a structured and repeatable way. Moving beyond resilience therefore depends on demonstrating adaptation, not merely disturbance. Future work must satisfy the formal classification criterion established in Section 2.5: simultaneous positivity of Jensen gain and post-disruption gain, followed by measurable susceptibility or recovery improvement under comparable perturbation.

Antifragility is therefore best understood not as a slogan, but as a disciplined research program. The present paper clarifies its definition, identifies measurable prerequisites, and outlines a feasible empirical pathway. The next step is to complete that pathway by introducing adaptation and testing whether systems improve under repeated bounded perturbation.

**Funding:** This research received no external funding.

**Data Availability Statement:** All data analyzed in this study are publicly available from third-party repositories. The CISSM Cyber Events Database, maintained by the Center for International and Security Studies at Maryland (University of Maryland), was the primary empirical dataset used for the sector-level analysis (16,729 events across 29 industry codes in the raw file at the time of access) and is available at https://cissm.umd.edu/cyber-events-database [19]. The HAI (HIL-Based Augmented Industrial Control System) Security Dataset, used as the process-control proxy in the OT/ICS analysis, is available at https://github.com/icsdataset/hai [21]. No new data were created by the authors during this study. Derived measures can be reproduced from the cited repositories using the procedures described in the manuscript.

**Conflicts of Interest:** The authors declare no conflicts of interest.

## ABBREVIATIONS

The following abbreviations are used in this manuscript:

| | |
|---|---|
| AFS | Antifragility Score |
| AFT | Theory of Antifragility |
| CI | Critical Infrastructure |
| CISSM | Center for International and Security Studies at Maryland |
| HAI | HIL-Based Augmented Industrial Control System (security dataset) |
| HIL | Hardware-in-the-Loop |
| IT | Information Technology |
| KEV | Known Exploited Vulnerabilities |
| NAICS | North American Industry Classification System |
| OT/ICS | Operational Technology / Industrial Control System |
| RMS | Root Mean Square |